\documentclass[12pt]{article}

\begin{document}

\title{LEVELS OF REALITY
AS SOURCE OF QUANTUM INDETERMINACY\thanks{Published in 
\textit{Determinismo e complessit\`a}, Fondazione Nova Spes and Armando Editore, 
Roma, 2000, pp. 127-158, edited by F. Tito Arecchi.}}

\date{}
\maketitle

\begin{center}
        Basarab NICOLESCU
\end{center}

\vspace{3mm}

\begin{center}  
        Laboratoire de Physique Nucl\'eaire et de
        Hautes Energies (LPNHE) \footnote{Unit\'e de Recherche des Universit\'es Paris 6 
        et Paris 7, Associ\'ee au CNRS.} \\
         LPTPE, Tour 12 E3 - 4, Place Jussieu 75252 Paris Cedex 05, France\\
                e-mail: nicolesc@in2p3.fr
\end{center}
\newpage
\section{Quantum physics and levels of Reality}
The major cultural impact of the quantum physics has certainly raised 
questions for the contemporary philosophical dogma of the existence 
of a single level of Reality \cite{BN85}.

Here the meaning we give to the word "Reality" is pragmatic and 
ontological at the same time.
 
By Reality I intend first of all to designate that which resists our 
experiences, representations, descriptions, images or mathematical 
formalizations. Quantum physics caused us to discover that 
abstraction is not simply an intermediary between us and Nature, a 
tool for describing reality, but rather, one of the constituent parts 
of Nature. In quantum physics, mathematical formalization is 
inseparable from experience. It resists in its own way by its 
simultaneous concern for internal consistency, and the need to 
integrate experimental data without destroying that self-consistency.

In so far as Nature participates in the being of the world one must 
ascribe an ontological dimension to the concept of Reality. Nature is 
an immense, inexhaustible source of the unknown which justifies the 
very existence of science. Reality is not only a social construction, 
the consensus of a collectivity, or an intersubjective agreement. It 
also has a trans-subjective dimension, to the extent that one simple 
experimental fact can ruin the most beautiful scientific theory.

By \textit{level of Reality } \cite{BN85} I intend to designate an ensemble of systems 
which are invariant under the action of certain general laws: for 
example, quantum entities are subordinate to quantum laws, which 
depart radically from the laws of the macrophysical world. That is to 
say that two levels of Reality are \textit{different} if, while passing from 
one to the other, there is a break in the laws and a break in 
fundamental concepts (like, for example, causality). No one has 
succeeded in finding a mathematical formalism which permits the 
rigorous passage from one world to another. Semantic glosses, 
tautological definitions or approximations are unable to replace a 
rigorous mathematical formalism. The recent decoherence models have 
nothing precise to say on the passage between the quantum level and 
the macrophysical level: in fact, the main problem is not decoherence 
but precisely \textit{coherence}.

There are even strong mathematical indications that the continuous 
passage from the quantum world to the macrophysical world would never 
be possible. But there is nothing catastrophic about this. The 
\textit{discontinuity} which is manifest in the quantum world is also manifest 
in the structure of the levels of Reality. That does not prevent the 
two worlds from co-existing.

The levels of Reality are radically different from the levels of 
organization as these have been defined in systemic approaches \cite{BN}. 
Levels of organization do not presuppose a break with fundamental 
concepts: several levels of organization appear at one and the same 
level of Reality. The levels of organization correspond to different 
structurings of the same fundamental laws. For example, Marxist 
economy and classical physics belong to one and the same level of 
Reality.
 
The emergence of at least two different levels of Reality in the 
study of natural systems is a major event in the history of knowledge.

The existence of different levels of Reality has been affirmed by 
different traditions and civilizations, but these affirmations were 
founded on religious dogma or on the exploration of the interior 
universe. 

In our century, in their questioning of the foundations of science, 
Edmund Husserl \cite{EH} and other scholars have discovered the existence 
of different levels of perception of Reality by the subject-observer. 
But these thinkers, pioneers in the exploration of a 
multi-dimensional and multi-referential reality, have been 
marginalized by academic philosophers and misunderstood by the 
majority of physicists, enclosed in their respective specializations.
The view I am expressing here is totally conform to the one of 
Heisenberg, Pauli and Bohr.
 
In fact, Werner Heisenberg came very near, in his philosophical 
writings, to the concept of "level of Reality". In his famous 
\textit{Manuscript of the year 1942} (published only in 1984) Heisenberg, who 
knew well Husserl, introduces the idea of three \textit{regions of reality}, 
able to give access to the concept of "reality" itself: the first 
region is that of classical physics, the second - of quantum physics, 
biology and psychic phenomena and the third - that of the religious, 
philosophical and artistic experiences \cite{WH}. This classification has a 
subtle ground: the closer and closer connectiveness between the 
Subject and the Object.

As we shall see in the following, the notion of levels of Reality 
will lead us to a general philosophical understanding of the nature 
of indeterminacy. If there was only one region or level of reality, 
it was impossible to conceive what means a true, irreducible 
indeterminacy, like the quantum one.

\section{The logic of the included middle}
Knowledge of the coexistence of the quantum world and the 
macrophysical world and the development of quantum physics has led, 
on the level of theory and scientific experiment, to the upheaval of 
what were formerly considered to be \textit{pairs of mutually exclusive 
contradictories} (A and non-A): wave \textit{and} corpuscle, continuity 
\textit{and} discontinuity, separability \textit{and} nonseparability, 
local causality \textit{and} global causality, symmetry \textit{and} 
breaking of symmetry, reversibility \textit{and} irreversibility of time, etc. 

The intellectual scandal provoked by quantum mechanics consists in 
the fact that the pairs of contradictories that it generates are 
actually mutually contradictory when they are analyzed through the 
interpretative filter of classical logic. This logic is founded on 
three axioms: 
\begin{enumerate}
        \item  \textit{The axiom of identity} : A is A.

        \item  \textit{The axiom of non-contradiction} : A is not non-A.

        \item  \textit{The axiom of the excluded middle} : There exists no 
        third term T which is at  the same time A and non-A.
\end{enumerate}
  
Under the assumption of the existence of a single level of Reality, 
the second and third axioms are obviously equivalent.

If one accepts the classical logic one immediately arrives at the 
conclusion that the pairs of contradictories advanced by quantum 
physics are mutually exclusive, because one cannot affirm the 
validity of a thing and its opposite at the same time: A \textit{and} non-A.

Since the definitive formulation of quantum mechanics around 1930 the 
founders of the new science have been acutely aware of the problem of 
formulating a new, "quantum logic." Subsequent to the work of 
Birkhoff and van Neumann a veritable flourishing of quantum logics 
was not long in coming \cite{TAB}. The aim of these new logics was to 
resolve the paradoxes which quantum mechanics had created and to 
attempt, to the extent possible, to arrive at a predictive power 
stronger than that afforded by classical logic. 

Most quantum logics have modified the second axiom of classical logic 
-- the axiom of non-contradiction -- by introducing non-contradiction 
with several truth values in place of the binary pair (A, non-A). 
These multivalent logics, whose status with respect to their 
predictive power remains controversial, have not taken into account 
one other possibility: the modification of the third axiom -- the 
axiom of the excluded middle.
 
History will credit St\'ephane Lupasco with having shown that the \textit{logic 
of the included middle} is a true logic, formalizable and formalized, 
multivalent (with three values: A, non-A, and T) and 
non-contradictory \cite{SL}. His philosophy, which takes quantum physics as 
its point of departure, has been marginalized by physicists and 
philosophers. Curiously, on the other hand, it has had a powerful 
albeit underground influence among psychologists, sociologists, 
artists, and historians of religions. Perhaps the absence of the 
notion of "levels of Reality" in his philosophy obscured its 
substance: many persons wrongly believed that Lupasco's logic 
violated the principle of non-contradiction. 

Our understanding of the axiom of the included middle -- \textit{there exists 
a third term T which is at the same time A and non-A}  -- is 
completely clarified once the notion of "levels of Reality" is 
introduced.
 
In order to obtain a clear image of the meaning of the included 
middle, we can represent the three terms of the new logic -- A, 
non-A, and T -- and the dynamics associated with them by a triangle 
in which one of the vertices is situated at one level of Reality and 
the two other vertices at another level of Reality. If one remains at 
a single level of Reality, all manifestation appears as a struggle 
between two contradictory elements (example: wave A and corpuscle 
non-A). The third dynamic, that of the T-state, is exercised at 
another level of Reality, where that which appears to be disunited 
(wave or corpuscle) is in fact united (quanton), and that which 
appears contradictory is perceived as non-contradictory.
 
It is the projection of T on one and the same level of Reality which 
produces the appearance of mutually exclusive, antagonistic pairs (A 
and non-A). A single level of Reality can only create antagonistic 
oppositions. It is inherently \textit{self-destructive} if it is completely 
separated from all the other levels of Reality. A third term, let us 
call it $T_{0}$, which is situated on the same level of Reality as that of 
the opposites A and non-A, can not accomplish their reconciliation.

The T-term is the key in understanding indeterminacy: being situated 
on a different level of Reality than A and non-A, it necessarily 
induces an \textit{influence} of its own level of Reality upon its 
neighbouring and different level of Reality: \textit{the laws of a given 
level are not self-sufficient to describe the phenomena occuring at 
the respective level}.

The entire difference between a triad of the included middle and an 
Hegelian triad is clarified by consideration of the role of \textit{time}. 
\textit{In a triad of the included middle the three terms coexist at the same 
moment in time}. On the contrary, each of the three terms of the 
Hegelian triad succeeds the former in time. This is why the Hegelian 
triad is incapable of accomplishing the reconciliation of opposites, 
whereas the triad of the included middle is capable of it. In the 
logic of the included middle the opposites are rather \textit{contradictories} 
: the tension between contradictories builds a unity which includes 
and goes beyond the sum of the two terms. The Hegelian triad would 
never explain the nature of indeterminacy.

One also sees the great dangers of misunderstanding engendered by the 
common enough confusion made between the axiom of the excluded middle 
and the axiom of non-contradiction . The logic of the included middle 
is non-contradictory in the sense that the axiom of non-contradiction 
is thoroughly respected, a condition which enlarges the notions of 
"true" and "false" in such a way that the rules of logical 
implication no longer concerning two terms (A and non-A) but three 
terms (A, non-A and T), co-existing at the same moment in time. This 
is a formal logic, just as any other formal logic: its rules are 
derived by means of a relatively simple mathematical formalism.
 
One can see why the logic of the included middle is not simply a 
metaphor, like some kind of arbitrary ornament for classical logic, 
which would permit adventurous incursions into the domain of 
complexity. \textit{The logic of the included middle is  the privileged logic 
of complexity}, privileged in the sense that it allows us to cross the 
different areas of knowledge in a coherent way, by enabling a new 
kind of simplicity.
 
The logic of the included middle does not abolish the logic of the 
excluded middle: it only constrains its sphere of validity. The logic 
of the excluded middle is certainly valid for relatively simple 
situations. On the contrary, the logic of the excluded middle is 
harmful in complex, transdisciplinary cases. For me, the problem of 
indeterminacy is precisely belonging to this class of cases.

\section{The G\"odelian unity of the world}
The transdisciplinary approach \cite{BN96} sets forth for consideration a 
multi-dimen\-sional Reality, structured by multiple levels replacing 
the single level of classical thought -- one-dimensional reality. 
This proposal is not enough, by itself, to justify a new vision of 
the world. We must first of all answer many questions in the most 
rigorous possible way. What is the nature of the theory which can 
describe the passage from one level of Reality to another? Is there 
truly a coherence, a unity of the totality of levels of Reality? What 
is the role of the subject-observer of Reality in the dynamics of the 
possible unity of all the levels of Reality? Is there a level of 
Reality which is privileged in relation to all other levels? What is 
the role of reason in the dynamics of the possible unity of 
knowledge? What is the predictive power of the new model of Reality 
in the sphere of reflection and action? Finally, is understanding of 
the present world possible ?
 
According to our model, Reality comprises a certain number of levels 
[1,2]. The considerations which follow do not depend on whether or 
not this number is finite or infinite. For the sake of clarity, let 
us suppose that this number is infinite.
 
Two adjacent levels are connected by the logic of the included middle 
in the sense that the T-state present at a certain level is connected 
to a pair of contradictories (A and non-A) at the immediately 
adjacent level. The T-state operates the unification of 
contradictories A and non-A but this unification is operated at a 
level \textit{different} from the one on which A and non-A are situated. The 
axiom of non-contradiction is thereby respected. Does this fact 
signify that we can obtain a complete theory, which will be able to 
account for all known and forthcoming results ?

There is certainly a coherence between different levels of Reality, 
at least in the natural world. In fact, an immense \textit{self-consistency} 
-- a cosmic bootstrap -- seems to govern the evolution of the 
universe, from the infinitely small to the infinitely large, from the 
infinitely brief to the infinitely long \cite{BN85}. A flow of information is 
transmitted in a coherent manner from one level of Reality to another 
level of Reality in our physical universe.

The logic of the included middle is capable of describing the 
coherence between the levels of Reality by an iterative process 
defined by the following stages: 1. A pair of contradictories (A, 
non-A) situated at a certain level of reality is unified by a T-state 
situated at a contiguous level of Reality; 2. In turn, this T-state 
is linked to a couple of contradictories (A', non-A'), situated at 
its own level; 3. The pair of contradictories (A', non-A') is, in its 
turn, unified by a T'-state situated at a different level of Reality, 
immediately contiguous to that where the ternary (A', non-A', T) is 
found. The iterative process continues indefinitely until all the 
levels of Reality, known or conceivable, are exhausted.
 
In other terms, the action of the logic of the included middle on the 
different levels of Reality induces an \textit{open}, \textit{G\"odelian}
structure of the unity of levels of Reality. This structure has considerable 
consequences for the theory of knowledge because it implies the 
impossibility of a complete theory, closed in upon itself.
 
In effect, in accordance with the axiom of non-contradiction, the 
T-state realizes the unification of a pair of contradictories (A, 
non-A) but it is associated, at the same time with another pair of 
contradictories (A', non-A'). This signifies that starting from a 
certain number of mutually exclusive pairs one can construct a new 
theory which eliminates contradictions at a certain level of Reality, 
but this theory is only temporary because it inevitably leads, under 
the joint pressure of theory and experience, to the discovery of new 
levels of contradictories, situated at a new level of Reality. In 
turn this theory will therefore be replaced by still more unified 
theories as new levels of Reality are discovered. This process will 
continue indefinitely without ever resulting in a completely unified 
theory. The axiom of non-contradiction is increasingly strengthened 
during this process. In this sense, without ever leading to an 
absolute non-contradiction, we can speak of an \textit{evolution of knowledge} 
which encompasses all the levels of Reality: knowledge which is 
forever open. Finer matter penetrates coarser matter, just as quantum 
matter penetrates macrophysical matter, but the reverse is not true. 
\textit{Degrees of materiality} induce an orienting arrow for tracing the 
transmission of information from one level to the other. This 
orienting arrow is associated with the discovery of more and more 
general, unifying, and encompassing laws.
 
The open structure of the unity of levels of Reality is in accord 
with one of the most important scientific results of the 20th century 
concerning arithmetic, the theorem of Kurt  G\"odel \cite{EN}.  G\"odel's 
theorem tells us that a sufficiently rich system of axioms inevitably 
lead to results which would be either undecidable or contradictory. 
The implications of  G\"odel's theorem have considerable importance for 
all modern theories of knowledge. First of all it does not only 
concern the field of arithmetic but also all mathematics which 
includes arithmetic. Now, obviously the mathematics which underlies 
theoretical physics include arithmetic. This means that all research 
for a complete physical theory is illusory.

In fact, the search for an axiomatic system leading to a complete 
theory (without undecidable or contradictory results) marks at once 
the apex and the starting point of the decline of classical thought. 
The axiomatic dream is unraveled by the verdict of the holy of holies 
of classical thought -- mathematical rigor.
 
The theorem that Kurt G\"odel demonstrated in 1931 sounded only a faint 
echo beyond a very limited circle of specialists. The difficulty and 
extreme subtlety of its demonstration explains why this theorem has 
taken a certain time to be understood within the mathematical 
community. Today, it has scarcely begun to penetrate the world of 
physicists. Wolfgang Pauli, one of the founders of quantum mechanics, 
was one of the first physicists to understand the extreme importance  
G\"odel's theorem has for the construction of physical theories \cite{WP}.

The  G\"odelian structure of the unity of levels of Reality associated 
with the logic of the included middle implies that it is impossible 
to construct a complete theory for describing the passage from one 
level to the other and, \textit{a fortiori}, for describing the unity of 
levels of Reality. 

If it does exist, the unity linking all the levels of Reality must 
necessarily be an open \textit{unity}.
 
To be sure, there is a coherence of the unity of levels of Reality, 
but we must remember that this coherence is \textit{oriented} : there is an 
arrow associated with all transmission of information from one level 
to the other. As a consequence of this, if coherence is limited only 
to the levels of Reality, it is stopped at the "highest" level and at 
the "lowest" level. If we wish to posit the idea of a coherence which 
continues beyond these two limited levels so that there is an open 
unity, one must conceive the unity of levels of Reality as a unity 
which is extended by a \textit{zone of non-resistance} to our experiences, 
representations, descriptions, images and mathematical 
formalizations. Within our model of Reality, this zone of 
non-resistance corresponds to the "veil" which Bernard d'Espagnat 
referred to as "the veil of the real" \cite{BE}. The "highest" level and 
the "lowest" level of the unity of levels of Reality are united 
across a zone of absolute transparence. But these two levels are 
different; from the point of view of our experiences, 
representations, descriptions, images, and mathematical 
formalizations, absolute transparence functions like a veil. In fact, 
the open unity of the world implies that that which is "below" is the 
same as that which is "above". The isomorphism between "above" and 
"below" is established by the zone of non-resistance. 

Quite simply, the non-resistance of this zone of absolute 
transparence is due to the limitations of our bodies and of our sense 
organs, limitations which apply regardless of the instruments of 
measure used to extend these sense organs. To claim that there is an 
infinite human knowledge (which excludes any zone of non-resistance), 
while simultaneously affirming the limitations of our body and our 
sense organs, seems to us a feat of linguistic sleight of hand. The 
zone of non-resistance corresponds to the sacred, \textit{that is to say to 
that which does not submit to any rationalization}.

The unity of levels of Reality and its complementary zone of 
non-resistance constitutes \textit{the transdisciplinary Object}.
 
A new \textit{Principle of Relativity } \cite{BN96} emerges from the coexistence 
between complex plurality and open unity : \textit{no one level of Reality 
constitutes a privileged place from which one is able to understand 
all the other levels of Reality}. A level of Reality is what it is 
because all the other levels exist at the same time. This Principle 
of Relativity is what originates a new perspective on religion, 
politics, art, education, and social life. In the transdisciplinary 
vision, Reality is not only multi-dimensional, it is also 
multi-referential. 

The different levels of Reality are accessible to human knowledge 
thanks to the existence of different \textit{levels of perception}, which are 
in bi-univocal correspondence with levels of Reality. These levels of 
perception permit an increasingly general, unifying, encompassing 
vision of Reality, without ever entirely exhausting it.
 
As in the case of levels of Reality the coherence of levels of 
perception presupposes a zone of non-resistance to perception.
 
The unity of levels of perception and its complementary zone of 
non-resistance constitutes \textit{the transdisciplinary Subject}.

The two zones of non-resistance of transdisciplinary Object and 
Subject must be \textit{identical} in order that the transdisciplinary Subject 
can communicate with the transdisciplinary Object. \textit{A flow of 
consciousness crossing the different levels of perception in a 
coherent manner must correspond to the flow of information crossing 
the different levels of Reality in a coherent manner}. The two flows 
are in a relation of \textit{isomorphism} thanks to the existence of one and 
the same zone of non-resistance. Knowledge is neither exterior nor 
interior: it is \textit{at the same time} exterior and interior. The study of 
the universe and the study of the human being sustain one another. 
The zone of non-resistance permits the unification of the 
transdisciplinary Subject and the transdisciplinary Object while 
preserving their difference.
 
Transdisciplinarity is the transgression of duality opposing binary 
pairs: subject/object,\hfill subjectivity/objectivity,\hfill 
matter/consciousness,\\
 nature/divine, simplicity/complexity, 
reductionism/holism, diversity/unity. This duality is transgressed by 
the open unity which encompasses both the universe and the human 
being.
 
The transdisciplinary model of Reality has, in particular, some 
important consequences in the study of \textit{complexity}. Without its 
contradictory pole of simplicity (or, more precisely, \textit{simplexity}) 
complexity appears as an increasingly enlarging \textit{distance} between the 
human being and Reality which introduces a self-destructive 
alienation of the human being who is plunged into the absurdity of 
destiny. The infinite simplicity of the transdisciplinary Subject 
corresponds to the infinite complexity of the transdisciplinary 
Object.
 
The Subject/Object problem was central for the founding-fathers of 
quantum mechanics. Pauli, Heisenberg and Pauli, as Husserl, Heidegger 
and Cassirer refuted the basic axiom of modern metaphysics: the 
clear-cut distinction between Subject and Object. Our considerations 
here are inscribed in the same framework.

\section{The death and the resurrection of Nature}
Modernity is particularly deadly. It has invented all kinds of 
"deaths" and "ends": the death of God, the death of Man, the end of 
ideologies, the end of history and, today, the end of science \cite{JH}.
 
But, there is a death which is spoken of much less, on account of 
shame or ignorance : \textit{the death of Nature}. In my view, this death of 
Nature is the source of all the other deadly concepts which were just 
invoked. In any case, the very word "Nature" has ended by 
disappearing from scientific vocabulary. Of course, the "man in the 
street", just as the scientist (in popularized works) still uses this 
word, but in a confused, sentimental way, reminiscent of magic.

Since the beginning of time we have not stopped modifying our vision 
of Nature \cite{RL}. Historians of science are in accord in saying that, 
despite all appearances to the contrary, there is not only one vision 
of Nature across time. What can there be in common between the Nature 
of so-called "primitive" peoples, the Nature of the Greeks, the 
Nature in the time of Galileo, of the Marquis de Sade, of Laplace or 
of Novalis? The vision of Nature of a given period depends on the 
imaginary which predominates during that period; in turn, that vision 
depends on a multiplicity of parameters: the degree of development of 
science and technology, social organization, art, religion, etc. Once 
formed, an image of Nature exercises an influence on all areas of 
knowledge. The passage from one vision to another is not progressive, 
continuous  -- it occurs by means of sharp, radical, discontinuous 
ruptures. Several contradictory visions can co-exist. The 
extraordinary diversity of visions of Nature explains why one cannot 
speak of Nature, but only of a certain nature in accord with the 
imaginary of a given period. 

The image of Nature has always had a multiform action: it has 
influenced not only science but also art, religion, and social life. 
This allows us to explain some strange synchronicities. Here I limit 
myself to but a single example: the simultaneous appearance of the 
theory of the end of history and of the end of science just before 
the beginning of the 3rd millenium. For example, unified theories in 
physics have as their aim the elaboration of a complete approach, 
founded on a unique interaction, which can predict everything (hence 
the name, "Theory of Everything"). It is quite obvious that if such a 
theory were formulated in the future, it would signify the end of 
fundamental physics, because there would be nothing left to look for. 
It is interesting to observe that both the idea of the end of history 
and of the end of science have simultaneously emerged from the "end 
of the century" imaginary.

Notwithstanding the abundant and fascinating diversity of images of 
Nature one can nevertheless distinguish three main stages: Magic 
Nature, Nature as Machine, and the Death of Nature. Magical thought 
views nature as a living organism, endowed with intelligence and 
consciousness. The fundamental postulate of magical thought is that 
of universal interdependence: Nature cannot be conceived outside of 
its relations with us. Everything is sign, trace, signature, symbol. 
Science, in the modern sense of this word, is superfluous.
 
At the other extreme, the mechanist and determinist thought of the 
18th and above all the 19th century (which, by the way, still 
predominates today) conceives Nature not as an organism, but as a 
machine. It suffices to disassemble this machine piece by piece in 
order to possess it entirely. The fundamental postulate of 
mechanistic and determinist thought is that Nature can be known and 
conquered by scientific methodology, defined in a way which is 
completely independent of human beings and separate from us.

The logical outcome of the mechanist and determinist vision is the 
Death of Nature, the disappearance of the concept of Nature from the 
scientific field. From the very beginning of the mechanistic vision, 
Nature as Machine, with or without the image of God as watchmaker, is 
split up into an ensemble of separate parts. From that moment on, 
there is no more need for a coherent whole, for a living organism, or 
even, for a machine which still kept the musty odor of finality. 
Nature is dead, but complexity remains. An astonishing complexity (in 
fact, often confused with "complication"), which penetrates each and 
every field of knowledge. But this complexity is perceived as an 
accident; we ourselves are considered to be an accident of complexity.

The Death of Nature is incompatible with a coherent interpretation of 
the results of contemporary science, in spite of the persistence of 
the neo-reductionistic attitude which accords exclusive importance to 
the fundamental building-blocks of matter and to the four known 
physical interactions. According to this neo-reductionist attitude, 
all recourse to Nature is superfluous and devoid of sense. In truth, 
Nature is dead only for a certain vision of the world  -- the 
classical vision. 

The rigid objectivity of classical thought is only viable in the 
classical world. The idea of total separation between an observer and 
a Reality assumed to be completely independent from that observer 
brings us to the verge of insurmountable paradoxes. In fact, a far 
more subtle notion of objectivity characterizes the quantum world: 
objectivity depends on the level of Reality in question.

Space-time itself no longer rests on a fixed concept. Our space-time 
which proceeds in four dimensions is not the only conceivable 
space-time. According to certain physical theories, it appears more 
an approximation, like a part of a space-time all the more rich for 
being the generator of possible phenomena. Supplementary dimensions 
are not the result of mere intellectual speculation. On the one hand, 
these dimensions are necessary to insure the self-consistency of the 
theory and the elimination of certain undesirable aspects. On the 
other hand, they do not have a purely formal character  -- they have 
physical consequences for our own scale. For example, according to 
certain cosmological theories, if the universe had been associated 
from the "beginning" of the big bang in a multi-dimensional 
space-time, supplementary dimensions would have remained forever 
hidden, unobservable; rather, their vestiges would be precisely the 
known physical interactions. By means of generalizing the example 
provided by particle physics, it becomes conceivable that certain 
levels of Reality correspond to a space-time different than that 
characterizing our own level. Moreover, complexity itself would 
depend on the nature of space-time as well.
 
We can make, like Heisenberg made \cite{WH}, a step further and assert that 
the classical four-dimensional space-time is, in fact, an 
\textit{anthropomorphic concept}, founded on our  sense-organs.

According to present scientific conceptions, \textit{matter} is far from being 
identical with substance. In the quantum world, matter is associated 
with a \textit{substance-energy-information-space-time complexus}.

It is somewhat mysterious why trajectories played such a central role 
in the formulation of modern physics. The quantum indeterminacy 
showed that trajectories are not a fondamental concept. In more 
recent years, a new discipline is born by the unexpected encounter 
between the theory of information and quantum mechanics: the Quantum 
Theory of Information \cite{DD}. This new-born science already poses a 
crucial question : are the \textit{information laws} more general, and 
therefore deeper, than the equations of movement? Are the central 
concepts of positions, speeds and trajectories of particles to be 
abandoned in favour of information laws which, in fact, could be 
valid not only for physics but also for other fields of knowledge? 
There were these last years fabulous experimental advances in the 
fields of non-separability, disentaglement, quantum cryptography and 
teleportation, in conjonction with the possible advent of quantum 
computers. This shows that notions like "levels of Reality" or 
"included middle" cease to be just theoretical speculations, by 
entering today  in the field of experiments and, tomorrow, in the 
everyday life.

We can assert that the notion itself of \textit{laws of Nature} completely 
changes its contents when compared with that of the classical vision. 
This situation can be summed up by three theses formulated by the 
well-known physicist Walter Thirring \cite{WT} :
 
1. \textit{The laws of any inferior level are not completely determined by 
the laws of a superior level}. Thus, notions well anchored in 
classical physics, like "fundamental" and "accidental," must be 
re-examined. That which is considered to be fundamental on one level 
can appear to be accidental on a superior level and that which is 
considered to be accidental or incomprehensible on a certain level 
can appear to be fundamental on a superior level.
 
2. \textit{The laws of an inferior level depend more on the circumstances of 
their emergence than on the laws of a superior level}. The laws of a 
certain level depend essentially on the local configuration to which 
these laws refer. There is therefore a kind of local autonomy of 
respective levels of Reality; however, certain internal ambiguities 
concerning laws of an inferior level of Reality are resolved by 
taking into account the laws of a superior level. It is the internal 
consistency of laws which reduces the ambiguity of laws.
 
3. \textit{The hierarchy of laws evolves at the same time as the universe 
itself}. In other words, the \textit{birth of laws} occurs simultaneously with 
the evolution of the universe. These laws pre-exist at the 
"beginning" of the universe as potentialities. It is the evolution of 
the universe which actualizes these laws and their hierarchy. A 
transdisciplinary model of Nature must integrate all this new 
knowledge of the emergent characteristics of the physical universe. 

The Thirring's description of the laws of Nature is in perfect 
agreement with our own considerations about the G\"odelian structure of 
Nature and knowledge. The problem of quantum indeterminacy can now be 
fully understood as \textit{the influence of the quantum level of Reality on 
our own macrophysical level of Reality}. Of course, the laws of the 
macrophysical level depend more, as Thirring writes, on "the 
circumstances of their emergence". From the point of view of the 
macrophysical level indeterminacy appears as accidental, 
incomprehensible, or at most as a rare event. But this reveals, in 
fact, an internal ambiguity which can be solved only by taking into 
account the laws of the quantum level. At this last level the 
indeterminacy is fundamental.

One can ask if one can not logically conceive a \textit{generalized 
indeterminacy}, which goes far beyond the problem of trajectories of 
particles. Heisenberg already considered the \textit{indeterminacy of 
language} \cite{WH} : the natural language can not express with arbitrary 
high precision all its elements, because the way of expressing acts 
in an essential manner on what is expressed. The indeterminacy of the 
natural language is just one example of the generalized indeterminacy 
generated by the G\"odelian structure of Nature and knowledge.

In conclusion, we can distinguish three major aspects of Nature in 
accordance with the transdisciplinary model of Reality :
 
(1) \textit{Objective Nature}, which is connected with the natural properties 
of the transdisciplinary Object; objective Nature is subject to 
subjective objectivity. This objectivity is subjective to the extent 
that the levels of Reality are connected to levels of perception. 
Nevertheless emphasis here is on objectivity, to the extent to which 
the methodology employed is that of science. 

2) \textit{Subjective Nature}, which is connected with the natural properties 
of the transdisciplinary Subject; subjective Nature is subject to 
objective subjectivity. This subjectivity is objective to the extent 
that the levels of perception are connected to levels of Reality. 
Nevertheless, emphasis here is on subjectivity, to the extent to 
which the methodology is employed is that of the ancient science of 
being, which crosses all the traditions and religions of the world.
 
3) \textit{Trans-Nature}, which is connected with a similarity in Nature  
which exists between the transdisciplinary Object and the 
transdisciplinary Subject. Trans-Nature concerns the domain of the 
sacred. It cannot be approached without considering the other two 
aspects of Nature at the same time. 

Transdisciplinary Nature has a ternary structure (objective Nature, 
subjective Nature, trans-Nature), which defines \textit{living Nature}. This 
Nature is living because it is there that life is present in all its 
degrees and because its study demands the integration of\textit{ lived 
experience}. The three aspects of Nature must be considered 
simultaneously in terms of their inter-relation and their conjunction 
within all the phenomena of living Nature \cite{LN}.

The study of living Nature asks for a new methodology -- 
transdisciplinary methodology \cite{BN96} -- which is different from the 
methodology of modern science and the methodology of the ancient 
science of being. It is the \textit{co-evolution} of the human being and of 
the universe which asks for a new methodology.

An attempt to elaborate a new \textit{Philosophy of Nature}, a privileged 
mediator of a dialogue between all the areas of knowledge, is one of 
the highest priorities of transdisciplinarity.

\vspace{2cm}


\end{document}